\documentclass[bibyear]{aa} 

\usepackage{graphicx}
\usepackage{amsmath}
\usepackage{float} 

\usepackage{txfonts}
\begin{document}

  \title{Updating the theoretical tidal evolution constants: Apsidal motion 
        and the moment of inertia }
 { }
   \subtitle{  }
\author{A. Claret\inst{1, 2}}
   \offprints{A. Claret, e-mail:claret@iaa.es. Tables 1-54  are   available 
in electronic form at the CDS via anonymous ftp. }
\institute{Instituto de Astrof\'{\i}sica de Andaluc\'{\i}a, CSIC, Apartado 3004,
            18080 Granada, Spain
        \and
     Dept. Física Teórica y del Cosmos, Universidad de Granada, Campus de Fuentenueva s/n, 10871 Granada, Spain
}
            \date{Received; accepted; }
\abstract
   {The theoretical apsidal motion constants are key tools to investigate 
        the stellar interiors in close eccentric binary systems. In addition, these constants   
        and the moment of inertia are also important to investigate  the tidal evolution of 
        close binary stars as well as of exo-planetary systems. }
   {The aim of the paper is to present new evolutionary models, based on the MESA 
        package, that include the internal structure constants (k$_2$,  k$_3$, and  k$_4$), 
         the radius of gyration, and the  gravitational potential energy for 
        configurations computed from the pre-main-sequence (PMS) up to the first ascent giant branch or beyond.  
        The calculations are available for the three metallicities [Fe/H]= 0.00, -0.50, and -1.00, which  
         take the recent investigations in less metallic environments into account. 
    This new set of models replaces the old ones, published about 15 years ago, using the code GRANADA.}
   {Core overshooting was taken into account using the mass-f$_{ov}$ relationship, which was  
        derived semi-empirically for models more massive  than 1.2 M$_{\odot}$. 
   The differential equations governing the apsidal motion constants, moment of inertia, 
   and the  gravitational potential energy were integrated simultaneously through a  
   fifth-order Runge-Kutta method with a tolerance level of 10$^{-7}$. }
   {The resulting models  (from 0.8 up to 35.0 M$_{\odot}$) are presented in 
    54 tables for the three metallicities,
        containing the usual 
        characteristics of an evolutionary model (age, initial masses, log T$_{\rm eff}$, log g, 
        and log L), the constants of internal structure  (k$_2$,  k$_3$, and  k$_4$), 
        the radius of gyration $\beta,$ and the factor $\alpha$ that is related with the 
        gravitational potential energy.  }
   {}

   \titlerunning {Tidal evolution }
   \maketitle
%

\section{Introduction}

Double-lined eclipsing binaries (DLEBS) are the main source of accurate, absolute 
dimensions of stars. In addition, some proximity effects can provide valuable 
information on the stellar interiors. In a close and eccentric binary system, the 
 stellar configurations are distorted by tides and rotation.  As a 
consequence of these distortions, there is a secular change of the position of 
the periastron. Such distortions can be described as a function of the internal 
structure of each star through the apsidal motion constants (k$_2$, k$_3,$ and k$_4$).  
For a complete description of the stellar distortions caused by tides and rotation, 
see Kopal (1959,1978).  
As the apsidal motion period depends strongly on the relative radius (r$^{-5}$), 
only systems with very accurate absolute dimensions can be used to perform the 
comparison between the theoretical and observed apsidal motion rates. 
An  advantage of the apsidal motion rates is that they  have been 
measured for a  wide range of masses. Therefore, one can investigate the 
interior of the stars  under very different 
physical conditions, thus   apsidal motion  is a good and complementary 
test for the theory of stellar evolution.

A  systematic effect was detected when the theoretical apsidal motion constants 
were  compared with the observed ones: the real stars seemed to be more mass 
concentrated than predicted by the  theoretical stellar models. For a short 
historical review on this point, see Claret\&Giménez (1993). These authors, 
using several improvements made to the stellar models available at that time 
(mainly opacities), were able to somewhat reduce
discrepancies for about 20 DLEBS. In 1997, analysing only  relativistic 
eclipsing binaries, Claret (1997) showed that general relativity predictions 
were able to explain  the shift in the periastron position 
without the need for an alternative theory of gravitation. The problematic 
case of DI Her  was solved observationally through the Rossiter-McLaughlin 
effect  some years later by Albrecht et al. (2009). A few months later, 
Claret and Torres\&Wolf (2009) redetermined the apsidal motion rate for DI Her  
using  new  times of minima. Also new evolutionary models were adopted that were
based on new absolute dimensions. With these data, these authors were 
able to reduce the difference between the theoretical and observed apsidal 
motion rate to only 10\%.

 Another small step forward was achieved by Claret\&Giménez (2010).  Using a 
 larger sample of  DLEBS (with improved absolute dimensions), they have shown 
 that the discrepancies between  the theoretical predictions and the 
 observational rates of apsidal motion have decreased significantly. 
 In that paper, in addition to the classical contribution to the apsidal 
 motion rate because of static tides, dynamic tides were also considered due to 
 the effects of the compressibility of the stellar  fluid, mainly  when the 
 system is  near the synchronism.

Since then, the accuracy of the absolute dimensions of the DLEBS have increased, which is 
the same for apsidal motion rates (Torres et al. 2010). The evolutionary 
stellar models have also improved (opacities, thermonuclear energy generation 
rates, equation of state, extra-mixing, etc). On the other hand,  the apsidal 
motion constants and the moment of inertia are also important in order to investigate 
the tidal evolution of close binary stars and exo-planetary systems 
(Hut 1981, 1982).  Therefore,  
a new generation of stellar models is necessary to be compared with the recent 
observational data about absolute dimensions and apsidal motion rates of DLEBS  
as well as with the data from the exo-planetary systems. Such updated evolutionary 
models are the main aim of the present research.  
The paper is organised as follows. Section 2 is dedicated to the introduction 
of the differential equations used to compute the theoretical apsidal motion constants 
as well as the moment of inertia and gravitational potential energy. Section 3 
is devoted to the analysis of the characteristics of the new models. Finally, Table 1   
summarises the properties of the models available  at the CDS (Centre de Donn\'ees 
Astronomiques de Strasbourg)  or those that come directly  from the author.    

\section {Stellar models, internal structure constants, and tidal evolution}

The stellar models were computed using  the Modules for Experiments in Stellar 
Astrophysics package (MESA; Paxton et al. 2011, 2013, 2015) version 7385, which does not include
the effects of  rotation. The adopted mixing-length parameter $\alpha_{MLT}$ was 
1.84 (the solar-calibrated value). However,  it is expected that the 
$\alpha_{MLT}$  parameter depends on  the evolutionary status and/or metallicity, 
according to the theoretical predictions of  the 3D simulations (Magic et al. 2015). 
Due to the different input physics of MESA, mainly 
the equation of state and opacities, a direct comparison of the  $\alpha_{MLT}$ adopted 
here  with those generated by 3D simulations is not straightforward.  In  all 
calculations, the  microscopic diffusion was included. For the opacities, the adopted  
element mixture is that by Asplund et al. (2009). The helium content is given by the 
following enrichment law: Y = 0.249 + 1.67 Z.  The covered mass range was 
from 0.8 up to  35.0 M$_{\odot}$.  For stars that are more massive  than 12 M$_{\odot}$, 
we adopted the scheme of mass loss from Vink et al. (2001)  with a multiplicative scale 
factor $\eta$ = 0.1. This is not to be confused with $\eta$ of the 
Radau equation (see below). For less massive stars in the red giants branch, we adopted   
the formalism by Reimers (1977) with $\eta$ = 0.2. For  AGB stars we adopted 
the formula given in Blocker (1995) with $\eta$ = 0.1. 
  In this paper,   convective core overshooting was introduced using the  diffusive 
approximation, given  by the free parameter f$_{ov}$ (Freytag et al. (1996)  and 
Herwig et al. (1997)).  In this approximation, the diffusion  coefficient in the 
overshooting region is given by the expression 
${D_{ov}} = D_o exp\left({{-2z\over{H_{\nu}}}}\right)$ and  D$_o$  is the diffusion 
coefficient at the convective boundary, $z$ is the geometric distance from the edge of 
the convective zone, H$_{\nu}$ is the velocity scale-height at the  convective boundary 
expressed as  H$_{\nu}$  = f$_{ov}$ H$_p$, and the coefficient  f$_{ov}$ is a free 
parameter governing the width of  the overshooting layer. It is known that models 
with core overshooting are more centrally concentrated in mass than their standard  
counterparts (Claret\&Giménez 1991).  At that time, it was usual to take the 
overshooting parameter ($\alpha_{ov}$) as a constant for the entire range 
of stellar masses. Instead, here, we adopted 
the relationship between the stellar mass and f$_{ov}$ found by Claret\&Torres  (2018)  
in their Fig. 3.  

Typically, DBLES that show apsidal motion are not very evolved. 
Nonetheless, all models were followed from the pre-main-sequence (PMS) up to 
the first ascent giant branch or beyond. The tables containing the models begin 
at the zero-age main-sequence (ZAMS), 
defined here  as the locus of the HR diagram  at which the
central hydrogen content drops to 99.4\% of its initial value.

 A significant observational advance   in apsidal motion studies 
 was recently achieved by Zasche\&Wolf (2019) who  investigated 21 eccentric eclipsing 
 binaries located in the small Magellanic Cloud. Taking this into account, we decided to 
 extend the calculations to the following three metallicities: [Fe/H] = 0.00, -0.50, and -1.00. 
 This  allows analysis of interpolating in the three grids in order to study DLEBS showing apsidal 
 motion and/or  tidal evolution that are also in  less  metallic environments.

The internal structure constants -  k$_2$, k$_3,$ and k$_4$ -  are derived by integrating 
the Radau equation

\begin{eqnarray}
{a{\rm d}\eta_{j}\over {\rm d}a}+ {6\rho(a)\over\overline\rho(a)}{(\eta_{j}\!+\!1)}+
{\eta_{j}(\eta_{j}\!-\!1)} = {j(j+1}), \, j=2,3,4
\end{eqnarray}

\noindent
where,

\begin{eqnarray}
\eta \equiv {{a}\over{\epsilon_{j}}} {{\rm d}\epsilon_{j}\over{{\rm d}a}}, 
\end{eqnarray}

\noindent
$a$ is the mean radius of the stellar configuration, $\epsilon_j$ is a  
measure of the deviation from sphericity, $\rho(a)$ is the 
mass density at the distance {\it a} from the centre, and $\overline\rho(a)$ 
is the mean mass density within a sphere of radius {\it a}.

The resulting apsidal motion constant of order $j$ is given by 

\begin{eqnarray}
k_{j} = {{j +1 - \eta_{j}(R)}\over{2\left(j+\eta_{j}(R)\right)}}
\end{eqnarray}

\noindent
where $R$ indicates the values  of $\eta_{j}$ at the surface of the star. 

We note that these  equations are  derived in the framework of static tides.  
For  the case of dynamic tides, more elaborated equations are necessary. 
 The rate of classical apsidal motion  was derived assuming that the orbital 
period is larger than the periods of the free oscillation modes. 
However, dynamic tides can significantly impact the theoretical 
predictions  based on static tides. This is  mainly due to the effects of the 
compressibility of the stellar fluid if the systems are near 
synchronism. In this case, for higher rotational angular velocities, 
additional deviations caused by resonances will appear if the forcing  
frequencies of the dynamic tides come into range of the free oscillation modes 
of the component stars. For numerical details on the impact of the dynamical 
tides  see Claret\&Willems (2002), Willems\&Claret (2003), and 
Claret\&Giménez (2010).

As previously mentioned, the present models were computed without  
taking rotation into account. To evaluate these effects, a correction on the internal 
structure constants  was introduced by Claret (1999) which is  given by 
$\Delta   {\rm log} k_2 \equiv {\rm log} k_{2, stan} - \lambda$, 
where  $\lambda=2V^{2}/(3gR)$,  $g$ is the surface gravity, and $V$ is the  equatorial  
rotational velocity.   

The effects of  general relativity  on the moment  of inertia and gravitational 
potential energy are not so important for stars at the  PMS, 
main-sequence (MS),  or for white dwarfs (WD). However,  for consistency with 
our previous papers on compact stars,  the 
relativistic formalism is adopted throughout. The moment of inertia 
is given by the following equation, where $\beta$ is the radius of gyration:     

\begin{eqnarray}
{J = {{8\pi}\over{3}}\int_{0}^{R} \Lambda(r)r^4\left[\rho'(r) +
        P(r)/c^2\right] dr }, \nonumber\\
I \approx  {J\over{\left(1 + {2GJ\over{R^3c^2}}\right)}} \equiv
{(\beta R)^2}M
\end{eqnarray}

\noindent
On the other hand, the amount of work  necessary to bring the entire spherical 
star in from infinity  is given by

\begin{eqnarray}
{\Omega = -4\pi\int_{0}^{R} {r^2\rho'(r)\left[\Lambda^{1/2}(r) - 1\right]dr} } 
\equiv -{\alpha} {G M^2\over{R}}, 
\end{eqnarray}

\noindent
where  $P(r)$ is the pressure,  $\rho'(r)$ the energy density, and the auxiliary
function, 
$\Lambda(r),$ is given by  $\left[1 - {2 G m(r)\over{r c^2}}\right]^{-1}$. 
 The parameter $\alpha$ is a dimensionless number that measures the relative mass concentration. 
For example, for polytropes, we have $\alpha = 3/(5-n)$, where $n$ is the polytropic index.

Equations 1, 4, 
and 5 were integrated simultaneously through a  fifth-order Runge-Kutta method, 
with a tolerance level of 10$^{-7}$.

The factors $\alpha$ and $\beta$  show very interesting  behaviour. Defining  
a  new function as 

\begin{eqnarray} 
\Gamma \equiv {\left[\alpha\beta\right]\over{\Lambda(R)^{0.8}}} 
\end{eqnarray} 

\noindent
Claret\&Hempel (2013) computed a series of stellar models  from  the PMS   up to the white dwarf  cooling sequences. 
They found a 
connection between the large variations of  $\Gamma$ during the intermediary 
evolutionary phases and the specific nuclear power. A threshold for the specific 
nuclear power was also found. Below this limit, $\Gamma$ is invariant 
($\approx 0.4$) and recovers the initial value at the  PMS. 

Other final products of the stellar evolution - neutron and quark stars 
- for which the effects of general relativity are strong, also present 
this invariance that is independent of the  mass and equation of state. 
Furthermore, by using a core-collapse supernova simulation, the PMS value 
of $\Gamma$ is recovered at the onset of the formation of a black hole. 
Therefore, regardless of the final products 
of the stellar evolution; white dwarfs; neutron, hybrid, and quark stars; or 
proto-neutron stars at the onset of the formation of a black hole, it can be concluded that they present 
a 'memory effect' and recover the fossil value of $\Gamma\approx$ 0.40 that is
acquired during the PMS (Claret 2014). The invariance of $\Gamma$ 
was also extended to another category of celestial bodies, the  gaseous 
planets, with masses ranging from 0.1 to 50 M$_J$ (Claret\&Hempel 2013). 
Such models were followed  from the gravitational contraction 
up to to  $\approx$ 20 Gyr. In addition, a
macroscopic stability criterion for neutron and quark star models, based on the 
properties of the relativistic product $\alpha\beta,$ was introduced.

The connection between the factors, $\alpha$ and $\beta,$ with the  Jacobi  
virial  theorem is clear for conservative systems recalling that such an 
equation can be written as 

\begin{eqnarray} 
{\partial^2\phi\over{\partial t^2}} = \Omega + 2 T = 2 E_T -\Omega. 
\end{eqnarray} 

In the above equation, the Jacobi function is given by 
$\phi = 1/2 {\sum_i}{m_\mathrm{i} (x_\mathrm{i}^2 + 
        y_\mathrm{i}^2 + z_\mathrm{i}^2)}$,  $ T$ is the kinetic energy and $E_T$ 
is the total energy. In the case of spherical symmetric  configurations,   
$\phi = {3/4 I} \equiv {3/4}M{(R\beta)}^2$. For conservative systems,  Eq. 7 can be 
transformed in a  differential equation with  only one
unknown function.  An important question coming from  the resulting 
equation is whether the product $\alpha\beta$  is constant for all stellar  evolutionary 
phases. However, as shown in the above mentioned papers, only during 
the PMS and the earlier stages of the main sequence is the product, $\alpha\beta,$
 almost constant. When  the stellar  models evolve out of the main sequence, this product 
increases during the later phases before reaching  the compact star stage. In this stage, the 
product recovers its original value ($\approx$0.40),  acquired at the PMS.  
 For  an insightful vision  of the virial theorem, see Collins (2003). 
 For a more detailed discussion on the relation of the  
Jacobi differential equation and stellar evolution, see Sect. 4 by Claret (2012).

\section {Tidal evolution constants: The apsidal motion  and the moment of inertia}

Returning to the main purpose of the paper (the constants of internal structure), 
we adopted some selected  models  by Claret (2004) for comparison purposes.  
The 2004 models were computed  for a slightly different chemical composition 
(X = 0.700, Z = 0.02, based on the Grevesse\&Sauval (1998)) mixture and  a 
different $\alpha_{MLT}$ = 1.68. 

 \begin{figure}
        \centering
        \includegraphics[,height=6.cm,width=8.cm]{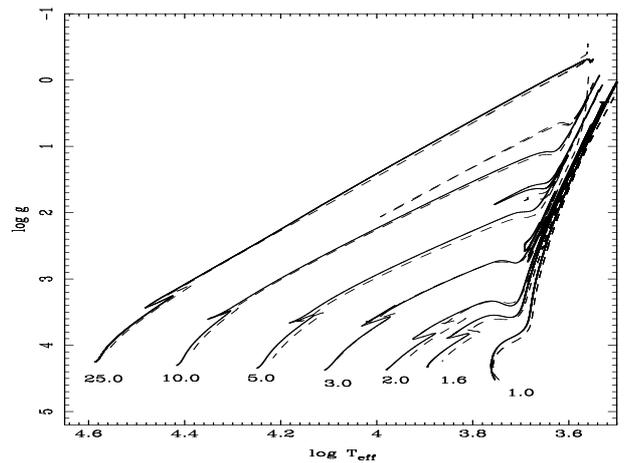}
        \caption{HR diagram for some selected models. The masses are indicated in solar units. 
        Continuous lines represent the present models while 
                dashed ones denote the models by Claret (2004). Solar composition.}
\end{figure}

\begin{figure}
        \centering
        \includegraphics[,height=6.cm,width=8.cm]{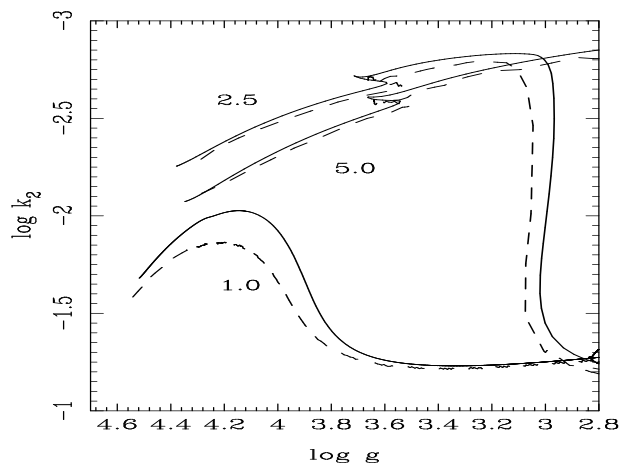}
        \caption{Apsidal motion constant of order two for new and old models.  
                The masses, in solar units, are indicated by  numbers. Same captions 
                as in  Fig. 1.}
\end{figure}

\begin{figure}
        \centering
        \includegraphics[,height=6.cm,width=8.cm]{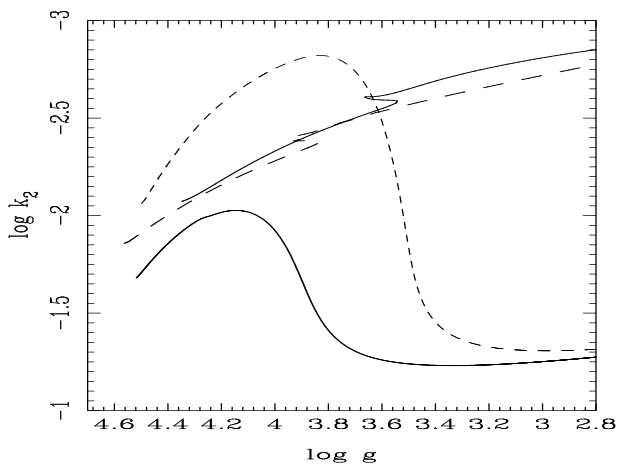}
        \caption{Apsidal motion constant of order two for present 
        models. Computed for [Fe/H] = 0.00 (continuous lines) and
                for [Fe/H]=-1.0 (dashed lines). Models with 1.00 and 5.00 M$_{\odot}$. }
\end{figure}

\begin{figure}
        \centering
        \includegraphics[,height=6.cm,width=8.cm]{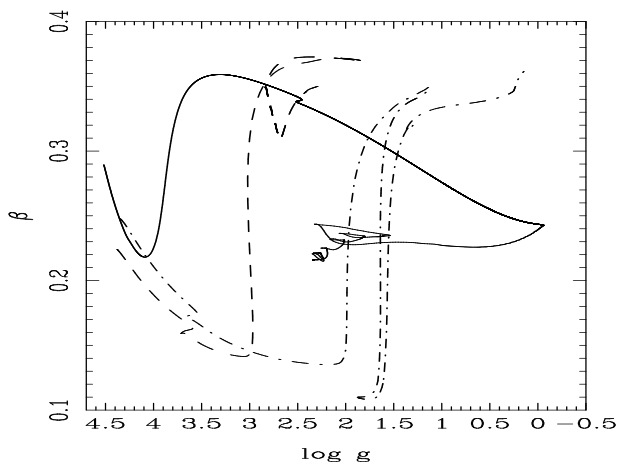}
        \caption{The evolution of radius of gyration for some selected models.
        Continuous line represents the model with 1.00 M$_{\odot}$, the dashed 
        and dashed-point lines denote the models with 2.5 and 5.0  
        M$_{\odot}$, respectively. Solar composition.}
\end{figure}

\begin{figure}
        \centering
        \includegraphics[,height=6.cm,width=8.cm]{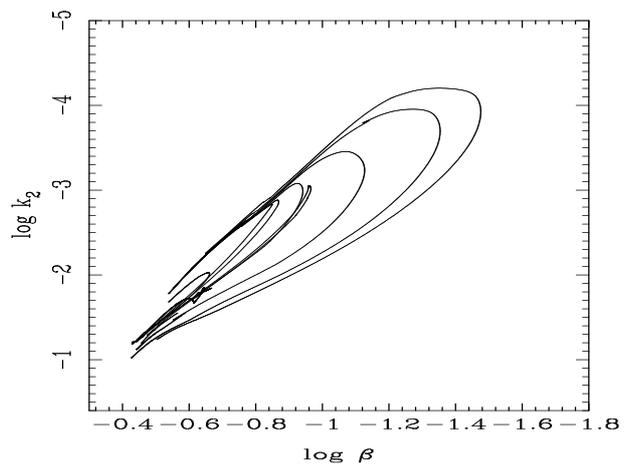}
        \caption{Relationship between evolving log $\beta$ and log k$_2$ for  models 
                with 1.00, 2.50, 5.00, 10.00, 15.00, 20.00, and 25.00  M$_{\odot}$. 
                Solar composition.}
\end{figure}

Figure 1 shows the HR diagram for the new and old models. There are two systematic 
effects. Firstly, the new models present higher effective temperatures than their 
old  counter-parts. Secondly, the radii of the new models are 
smaller than the old ones, and the effect is smaller for low-mass stars.

Concerning the theoretical apsidal motion constants, in general, the new models 
are more concentrated in mass than the old ones (Fig. 2). However, the differences 
are not constant and progressively decrease from less massive to more massive models. 
Considering, for example,  models of 1.00 M$_{\odot}$, the difference is of the 
order of 0.20 dex (in the logarithmic scale) in the middle of the main sequence, 
which is of the order of the mean observational errors. For the more massive 
models (e.g. 2.50 and 5.00 M$_{\odot}$), the differences between the new models 
and the old  models are only of the order of 0.05 dex, though they are still important.

The differences between the two sets of models discussed above have diverse  
causes, such as the adopted  values of $\alpha_ {MLT}$  (1.84 and 1.68). The 
equation of state also contributes. It is important to note that this is mainly the case for the less massive models. 
 The rates of mass loss, which are different from those adopted in the 2004 models,  
also influence both the shape of the tracks in the HR diagram and their internal structure. 
Another reason for the differences in the 2004 models is that a constant value for the amount of core overshooting was adopted for all masses; conversely, as previously mentioned, we introduced a dependence on the stellar mass for the core overshooting as given by Claret\&Torres (2018).

The effects of the metallicity on the internal structure is shown in Fig. 3 
for models with 1.00 and 5.00 M$_{\odot}$ for [Fe/H=0.00] with continuous lines,  and 
for [Fe/H=-1.00] with dashed lines.  The differences  
between the [Fe/H=0.00] and [Fe/H=-1.00] models are important. This mainly concerns less 
massive models during the main sequence and can achieve up to 0.80 dex for the 
1.00 M$_{\odot}$ models. These differences, in general, decrease with 
the stellar mass.

For homogeneous models, the radius of gyration shows a different behaviour  
for masses larger or smaller than  $\approx$1.5 M$_{\odot}$, that is there is a 
minimum of $\beta$ around this value. Such behaviour is 
connected with the change in the dominant source of thermonuclear energy 
from the  pp chain to the CNO cycle (Fig. 2 by Claret\&Giménez 1989). In Fig. 4,  
we show the evolution of $\beta$ for 
models of 1.00, 2.50, and 5.00  M$_{\odot}$. 
 When the star evolves, the behaviour of $\beta$ becomes more complex, showing 
 some loops for advanced evolutionary phases. Despite the complex behaviour of 
 the apsidal motion constant of order two and the radius of gyration as a function 
 of log g shown in Figs. 2 and 4, both have an approximately linear relationship  that is
 independent of their evolution for a wide range of stellar masses. The  
 resulting lobe shown in Fig. 5 is due to nuclear reactions and  chemical 
 inhomogeneities which lead 
 to structural changes.   The aspect of Fig. 5 is not surprising given that 
 log k$_2$ and log $\beta$  give a measure of the mass concentration for each 
 evolutionary stage of the models through the integration of the respective 
 differential equations of second and first order, respectively.

Finally, Table 1 summarises the  data available at the CDS or directly from the author. 
Additional calculations for other chemical compositions and/or for specific 
masses  can be performed upon request.

\begin{acknowledgements} 
 I am grateful to the anonymous referee for the careful reading of the paper 
and for her/his helpful suggestions. 
I am grateful to  B. Rufino, G. Torres, and V. Costa for their useful comments.         
The Spanish MEC (AYA2015-71718-R and ESP2017-87676-C5-2-R) is gratefully 
acknowledged for its support during the development of this work. 
AC also acknowledges financial support from the State Agency for Research of 
the Spanish MCIU through the Center of Excellence Severo Ochoa award for 
the Instituto de Astrofísica de Andalucía (SEV-2017-0709).
 This research has made use of the SIMBAD database, operated at the CDS, 
 Strasbourg, France, and of NASA's Astrophysics Data System Abstract Service.
\end{acknowledgements}

{}

\begin{table}
        \caption[]{Stellar models}
        \begin{flushleft}
                \begin{tabular}{lcccl}                         
                        \hline                         
                        Name    & Initial mass (solar mass)   &  [Fe/H]  \\ 
                        \hline   
                        Table2   &0.80  & 0.00\\
                        Table3   &1.00  & 0.00\\
                        Table4   &1.20  & 0.00\\
                        Table5   &1.40  & 0.00\\
                        Table6   &1.60  & 0.00\\
                        Table7   &1.80  & 0.00\\
                        Table8   &2.00  & 0.00\\
                        Table9   &2.50  & 0.00\\
                        Table10  &3.00  & 0.00\\
                        Table11  &5.00  & 0.00\\
                        Table12  &7.00  & 0.00\\
                        Table13  &10.00 & 0.00\\
                        Table14  &12.00 & 0.00\\
                        Table15  &15.00 & 0.00\\
                        Table16  &20.00 & 0.00\\
                        Table17  &25.00 & 0.00\\
            Table18  &30.00 & 0.00\\            
                        Table19  &35.00 & 0.00\\
                        Table20  &0.80  & -0.50\\
                        Table21  &1.00  & -0.50\\
                        Table22  &1.20  & -0.50\\
                        Table23  &1.40  & -0.50\\
                        Table24  &1.60  & -0.50\\
                        Table25  &1.80  & -0.50\\
                        Table26  &2.00  & -0.50\\
                        Table27  &2.50  & -0.50\\
                        Table28  &3.00  & -0.50\\
                        Table29  &5.00  & -0.50\\
                        Table30  &7.00  & -0.50\\
                        Table31  &10.00 & -0.50\\
                        Table32  &12.00 & -0.50\\
                        Table33  &15.00 & -0.50\\
                        Table34  &20.00 & -0.50\\
                        Table35  &25.00 & -0.50\\
            Table36  &30.00 & -0.50\\           
                        Table37  &35.00 & -0.50\\
                        Table38  &0.80  & -1.00\\
                        Table39  &1.00  & -1.00\\
                        Table40  &1.20  & -1.00\\
                        Table41  &1.40  & -1.00\\
                        Table42  &1.60  & -1.00\\
                        Table43  &1.80  & -1.00\\
                        Table44  &2.00  & -1.00\\
                        Table45  &2.50  & -1.00\\
                        Table46  &3.00  & -1.00\\
                        Table47  &5.00  & -1.00\\
                        Table48  &7.00  & -1.00\\
                        Table49  &10.00 & -1.00\\
                        Table50  &12.00 & -1.00\\
                        Table51  &15.00 & -1.00\\
                        Table52  &20.00 & -1.00\\
                        Table53  &25.00 & -1.00\\
            Table54  &30.00 & -1.00\\           
                        Table55  &35.00 & -1.00\\
                        \hline
                        \hline
                \end{tabular}
        \end{flushleft}
\end{table}


\begin{thebibliography} {}
        
\bibitem{} Albrecht, S., Reffert, S., Snellen, I. A. G., 
\& Winn, J. N. 2009, Nature, 461, 373

\bibitem{} Asplund, M., Grevesse, N., Sauval, A. J., 
\& Scott, P. 2009, ARA\&A, 47, 481

\bibitem{} Blocker, T. 1995, A\&A 297, 727

\bibitem{} Claret, A. 1997, A\&A, 327, 11C

\bibitem{} Claret, A.  1999, A\&A, 350, 56

\bibitem{} Claret, A.,  2004, A\&A, 424, 919

\bibitem{} Claret, A.,  2012, A\&A, 543, A67

\bibitem{} Claret, A.,  2014, A\&A, 562A, 31C

\bibitem{} Claret, A., Hempel, M. 2013, A\&A, 552A, 29C

\bibitem{} Claret, A., Gim\'enez, A. 1989, A\&AS, 81, 37

\bibitem{} Claret, A., Gim\'enez, A. 1991, A\&A, 244, 319

\bibitem{} Claret, A., Gim\'enez, A. 1993, A\&A, 277, 487

\bibitem{} Claret, A., Gim\'enez, A. 2010, A\&A, 519A, 57C

\bibitem{} Claret, A., Torres, G. Wolf, M.  2010, A\&A, 515, A4

\bibitem{} Claret, A.,  Torres G. 2018, ApJ, 859, 100 

\bibitem{} Claret, A., Willems, B. 2002, A\&A, 388, 518

\bibitem{} Collins II, G. W. 2003, The Virial Theorem in Stellar 
Astrophysics, Web Edition

\bibitem{} Freytag, B., Ludwig, H.-G.,  Steffen, M. 1996, A\&A, 313, 497

\bibitem{} Grevesse, N., Sauval, A. J. 1998, SSRv, 85, 161 

\bibitem{} Herwig, F., Bloecker, T., Schoenberner, D., El Eid, M. 1997, 
A\&A,324, L81

\bibitem{} Hut, P. 1981, A\&A, 99, 126

\bibitem{} Hut, P. 1982, A\&A, 102, 37H

\bibitem{} Kopal, Z. 1959,  Close Binary Systems, Chapman and Hall, London

\bibitem{} Kopal, Z. 1978,  Dynamics of Close Binary Systems, 
Reidel, Dordrecht, Holland

\bibitem{} Magic, Z., Weiss, A.,  Asplund, M. 2015, A\&A, 573, A89

\bibitem{} Paxton, B., Bildsten, L., Dotter, A., et al. 2011, ApJS, 192, 3

\bibitem{} Paxton, B., Cantiello, M., Arras, P., et al. 2013, ApJS, 208, 4

\bibitem{} Paxton, B., Marchant, P., Schwab, J., et al. 2015, ApJS, 220, 15

\bibitem{} Reimers, D. 1977, A\&A, 61, 217

\bibitem{} Torres, G., Andersen, J., Gim\'enez, A. 2010 A\&AR, 18, 67 

\bibitem{} Vink, J.S., de Koter, A., Lamers, H.J.G.L.M., 2001, A\&A, 369, 574

\bibitem{} Willems, B., Claret, A.  2003, A\&A, 410, 289

\bibitem{} Zasche, P.,  Wolf, M. 2019, AJ, 157, 87Z

\end{thebibliography}
\end{document}